\documentclass[twocolumn,showpacs,amsmath,amssymb,nofootinbib,prl]{revtex4}

\usepackage{textcomp}
\usepackage{graphicx}
\usepackage{bm}
\usepackage[all,2cell]{xy}
\usepackage{color}

\newcommand{\be}{\begin{equation}}
\newcommand{\ee}{\end{equation}}
\newcommand{\bea}{\begin{eqnarray}}
\newcommand{\ba}{\begin{array}}
\newcommand{\eea}{\end{eqnarray}}
\newcommand{\bes}{\begin{subequations}\bea}
\newcommand{\ees}{\eea\end{subequations}}
\newcommand{\ea}{\end{array}}
\newcommand{\bs}[1] {\boldsymbol{#1}}
\newcommand{\tc}[1]{\textcolor{black}{#1}}
\newcommand{\tcb}[1]{\textcolor{black}{#1}}
\newcommand{\f}{\varphi}
\newcommand{\C}{\zeta}
\newcommand{\ALPHA}{a}
\newcommand{\BETA}{b}
\newcommand{\GAMMA}{c}
\newcommand{\CHI}{h}

\begin{document}

\title{\tcb{Efficiency statistics at all times: Carnot limit at finite power}}

% \title{\tcb{Probability distribution of the efficiency at finite time: Carnot limit at nonvanishing power output}}

\author{M. Polettini}
\email{matteo.polettini@uni.lu}
\affiliation{
Complex Systems and Statistical Mechanics, Physics and Materials Research Unit, University of Luxembourg, 162a avenue de la Fa\"iencerie, L-1511 Luxembourg (G. D. Luxembourg)} 

\author{G. Verley}
\email{gatien.verley@th.u-psud.fr}
\affiliation{
Complex Systems and Statistical Mechanics, Physics and Materials Research Unit, University of Luxembourg, 162a avenue de la Fa\"iencerie, L-1511 Luxembourg (G. D. Luxembourg)} 

\author{M. Esposito}
\email{massimilano.esposito@uni.lu}
\affiliation{
Complex Systems and Statistical Mechanics, Physics and Materials Research Unit, University of Luxembourg, 162a avenue de la Fa\"iencerie, L-1511 Luxembourg (G. D. Luxembourg)}

\begin{abstract}
% Keywords: Finite time behavior, criticality at tight coupling, fluctuation theorem, system/environment duality.

We derive the statistics of the efficiency under the assumption that thermodynamic fluxes fluctuate with normal law, parametrizing it in terms of time, macroscopic efficiency, and a coupling parameter $\C$. It has % related to the figure of merit of the machine,
a peculiar behavior: No moments, one sub- and one super-Carnot maxima corresponding to reverse operating regimes (engine/pump), the most probable efficiency decreasing in time. %, and a minimum moving towards the least-likely Carnot efficiency.
The limit $\C\to 0$ where the Carnot bound can be saturated gives rise to two extreme situations, one where the machine works at its macroscopic efficiency, with Carnot limit corresponding to no entropy production, and one where for a transient time scaling like $1/\C$ microscopic fluctuations are enhanced in such a way that the most probable efficiency approaches Carnot at finite entropy production. % The latter situation, invisible to the macroscopic theory, might be encountered at dynamical phase transitions, giving an indication on how to devise efficient machines with finite power output.
\end{abstract} 


% 88.05.Bc energy efficiency 
% 05.70.Ln Nonequilibrium Thermodynamics
% 05.70.Fh Phase transitions in in statistical mechanics and thermodynamics, 

\maketitle

Efficiency quantifies how worth a local gain at the expense of  a global loss is. In thermodynamics, ``losses'' are measured by the rate $\bar{\sigma}_2 > 0$ at which entropy is externalized to the environment in the form of a degraded form of energy, while ``gain'' is the rate $-\bar{\sigma}_1$ at which entropy is expelled from a system to upgrade its own state. Globally, entropy is produced at rate $\bar{\sigma} = \bar{\sigma}_2 + \bar{\sigma}_1$, and the second law of thermodynamics $\bar{\sigma} \geq 0$ conveys that locally one cannot earn more of what is globally lost. Then, the efficiency $\bar{\eta} = - \bar{\sigma}_1/\bar{\sigma}_2$ is bounded by the (scaled) Carnot efficiency $\eta_c= 1$. Alas, the craving of this limit is deluded by the fact it occurs at zero power, which is useless for any activity to be accomplished in a reasonable time.

This picture is only tenable for macroscopic systems. For microscopic systems subject to random fluctuations, the concept of a stochastic efficiency has been recently introduced by Verley et al. \cite{gatien1,gatien2}. The first notion one has to revise is that a fluctuating efficiency can indeed exceed the Carnot limit, when in a machine designed to convert in average a form of input power into a form of output power (e.g. an engine producing work at the expense of a heat flow), for a rare event the input and output are reversed (e.g. a pump that employs mechanical work to absorb heat). Moreover, it has been observed that for time-symmetric protocols in the long time limit the Carnot efficiency becomes the least probable in a ``large deviation'' sense \cite{ld} -- a very counterintuitive and fascinating result  \tcb{that, in its time-asymmetric variant \cite{gatien2,gingrich}, is already subject to experimental inquiry \cite{roldan}. Corrections at long finite times have been estimated in Ref.\,\cite{gingrich}.}

In this work we derive the full probabilit density function (p.d.f.) of the efficiency, under the assumption that thermodynamic fluxes are distributed with multivariate Gaussian with cumulants growing linearly in time. The efficiency p.d.f. displays quite peculiar features. In particular, it does not afford moments of any order, so that there is no average efficiency and mean-square error. Experimentally, this implies that any data analysis should focus on most probable values. About the latter, after an initial transient the distribution becomes bimodal,  as observed numerically in Ref.\,\cite{single}. As time elapses, the more pronounced maximum drifts towards the always smaller macroscopic value of the efficiency, while a less pronounced maximum at higher efficiency moves in the super-Carnot region towards infinity. We provide a clear physical interpretation of these two peaks. 
Finally, we argue that the macroscopic framework fails to capture another way of approaching Carnot at finite entropy production, at finite time, when microscopic fluctuations are enhanced so to affect the macroscopic behavior.

Macroscopic nonequilibrium thermodynamics \cite{degroot} is rooted on two assumptions, both of which are today being challenged in the framework of the stochastic theory of nonequilibrium thermodynamics \cite{st1,st3}:
Certain fluxes  $\bs{x} = (x_1,x_2)$, with units of an extensive physical quantity per time, take definite values $\bar{\bs{x}}$; Fluxes are linearly related to their conjugate thermodynamic forces $\bs{f}$ via $\bar{\bs{x}} = L \bs{f}$, where the linear response matrix $L$ is assumed to be positive-semidefinite and symmetric by virtue of the Onsager reciprocity relations, yielding a non-negative macroscopic entropy production rate $\bar{\sigma} = \bs{f} \cdot L \bs{f}$.

We relax the first assumption, by supposing that at a given time $t$ fluxes $\bs{x}$ are distributed with law $P_t(\bs{x})$. Each current produces entropy at rate $\sigma_i = f_i x_i$, for a total entropy production rate $\sigma = \sigma_1+\sigma_2$, with units of $k_B$ per time. Then, the adimensional efficiency
\be
\eta = - \frac{f_1 x_1}{f_2 x_2} = - \frac{\sigma_1}{\sigma_2}
\ee
is a stochastic variable distributed with p.d.f.
\bea
P_t(\eta) & = & \int dx_1 dx_2 \, \delta\left(\eta + \frac{x_1 f_1}{x_2f_2}\right) P_t(x_1,x_2) \nonumber \\ 
& = & \f \int dx |x| P_t\left(-\f \eta  \, x,x\right), \label{eq:effpdf}
\eea
where $\f = f_2/f_1$ \tc{can be assumed to be positive}. A remarkable fact one immediately encounters is that the efficiency can fluctuate beyond the Carnot limit. The probability of an efficiency higher than Carnot coincides with the probability of negative entropy production rate,
\bes
P_t(\eta < 1) & = & P_t({\sigma > 0}) = \big\langle \theta(\sigma) \big\rangle_t  \\  
P_t(\eta > 1) & = & P_t({\sigma < 0}) = \big\langle \theta(\sigma) e^{-t\sigma} \big\rangle_t,
\ees
where $\theta$ is Heaviside's step function. The rightmost equations follow \tcb{from} the fluctuation theorem \cite{ft,polespo} 
\be
\frac{P_t(\sigma)}{P_t(-\sigma)} = e^{t \sigma}, \label{eq:ft}
\ee
which states that processes producing negative entropy are exponentially disfavored  with respect to those producing positive entropy. Therefore, that super-Carnot efficiencies are unlikely compared to sub-Carnot efficiencies is an incarnation of the fluctuation theorem.

Exact results can be obtained by assuming that fluxes are distributed with normal multivariate density function
\be
P_t(\bs{x}) = \frac{t}{4\pi \sqrt{|L|}} \exp \left[ - {\frac{t}{4} (\bs{x}-\bar{\bs{x}})\cdot  L^{-1} (\bs{x}-\bar{\bs{x}})} \right], \label{eq:normal}
\ee
where $|\cdot|$ is the determinant. That (one-half) the correlation matrix  should be identified with the linear response matrix is corroborated by the Green-Kubo relations 
\be
L_{ij} = \frac{t}{2} \langle (x_i - \bar{x}_i)(x_j-\bar{x}_j) \rangle, \label{eq:gk}
\ee
another well-known consequence of the fluctuation theorem \cite{andrieux}. 
The time dependence in Eq.\,(\ref{eq:normal}) is due to the fact that the time-integrated fluxes $t\bar{\bs{x}}$ increase linearly in time, and correspondingly so do their cumulants. Under these assumptions the efficiency p.d.f. Eq.\,(\ref{eq:effpdf}) can be exactly calculated \cite{supp}. It only depends on four adimensional parameters: The macroscopic efficiency $\bar{\eta}$, the coupling parameter $\C = |L|/(L_{11} L_{22})\in [0,1]$ that for thermoelectric devices \cite{benenti} is related to the so-called figure of merit $zT = 1/\C-1$, the average entropy production rate $\bar{\sigma}$, which sets the time scale and can be reabsorbed by a time reparametrization $\tau = t\bar{\sigma}$, and $\epsilon=\pm 1$. Being $\bar{\sigma}$ the only extensive parameter, large $\tau$ \tc{stands both for large times and the macroscopic limit}. We obtain \cite{supp}
\be
P_\tau(\eta) = \frac{ e^{- \frac{\tau}{4} } }{\pi \ALPHA(\eta)\sqrt{|C|}} \left\{1 + \sqrt{\pi \tau} \, \CHI(\eta) \, e^{\tau \CHI(\eta)^2} \mathrm{\,erf}\,\left[\sqrt{\tau} \CHI(\eta)\right] \right\} \label{eq:pdf}
\ee
where $\mathrm{erf}$ is the error function and 
\begin{subequations}\label{eq:sub}
\bea
\ALPHA(\eta) & = &
(1-\eta)^2+ \frac{1}{|C|}\left(\frac{\eta - \bar{\eta}}{1 - \bar{\eta}}\right)^2 \\
\CHI(\eta) & = &  \frac{1-\eta}{2\sqrt{\ALPHA(\eta)}}. 
\ees 
Here, $|C| = |L|f_1^2f_2^2/\bar{\sigma}^2$ is the determinant of the matrix with dimensionless entries $C_{ij} = L_{ij} f_i f_j/\bar{\sigma}$. It can be expressed in terms of our parameters as
\be
|C| =  \frac{zT}{2} \left(1  +\epsilon \sqrt{1 - \frac{4}{zT} \frac{\bar{\eta}}{(1-\bar{\eta})^2} } \right) - \frac{\bar{\eta}}{(1-\bar{\eta})^2}, \label{eq:C}
\ee
where $\epsilon = \pm$ accounts for the existence of two probability distributions corresponding to given parameters. For $|L|$ to be real, the known bound
\be
\bar{\eta} \leq \frac{1- \sqrt{\C}}{1+ \sqrt{\C}}  \label{eq:limit}
\ee
must hold  \cite{benenti}. Importantly, $\ALPHA(\eta)$ is positive semidefinite. 

\begin{figure}
  \centering
 \includegraphics[width=240pt]{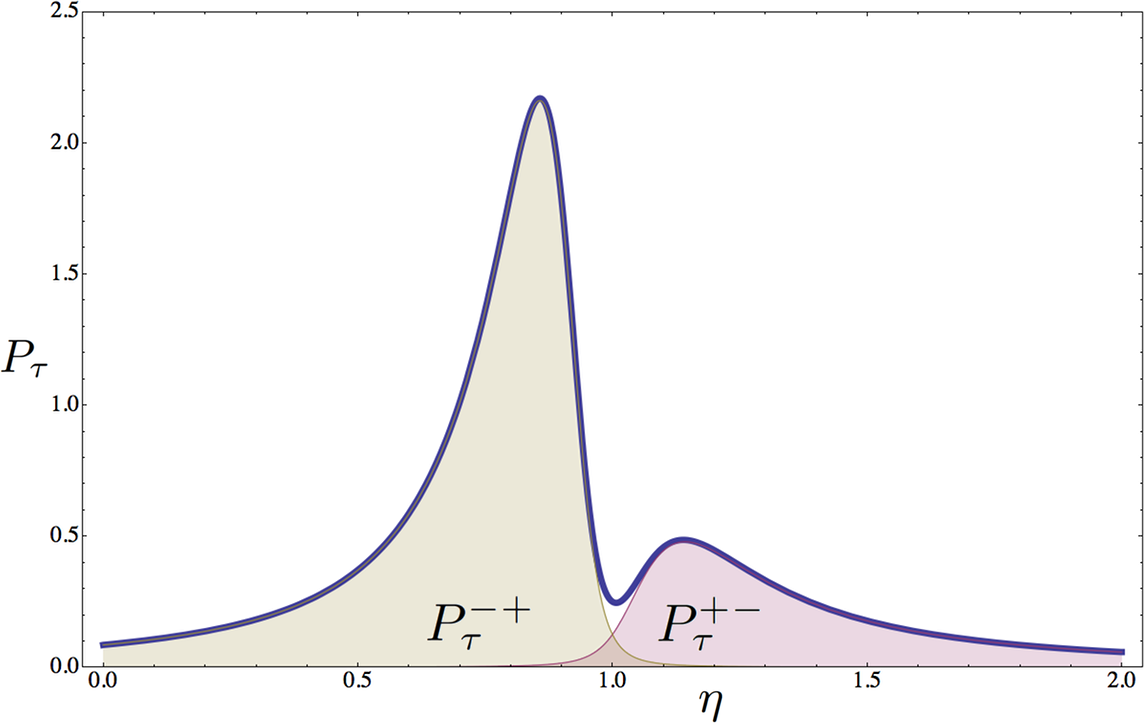}
  \caption{ \label{distribution} \tcb{Bold curve: efficiency distribution $P_\tau(\eta)$} for parameter values $\C = 0.01$, $\bar{\eta} = 0.6$, $\tau = 10$, $\epsilon = +1$. Filled curves beneath: $P^{-+}_{\tau}(\eta)$ and $P^{+-}_{\tau}(\eta)$, showing that  each maximum is mostly due to one working mode of the engine.}
\end{figure}
Let us study the efficiency p.d.f. in detail. First, it is a power-law distribution with tails
\be
P_\tau(\eta \to \pm \infty) \propto  \eta^{-2},
\ee
\tcb{which, after submission of this letter, has been proven to be a universal property of efficiency distributions \cite{proesmans}.} As a consequence, it does not afford finite moments of any order. Hence, the macroscopic efficiency $\bar{\eta}$ is {\it not} the average efficiency $\langle \eta \rangle_t$, which is not finite.

In Fig.\,\ref{distribution} the efficiency distribution is plotted as the bold curve. Remarkably, for a large class of parameters it displays two maxima at $\eta_m,\eta_m^\ast$ and a minimum, the latter slightly off the Carnot efficiency. Hence, not only super-Carnot efficiencies are possible, but indeed there appears a local maximum with efficiency higher than Carnot. To understand its physical origin, we distinguish four operational regimes of the machine, according to the signs of the two contributions $\sigma_1$ and $\sigma_2$ to the entropy production rate. \tc{The two regimes contributing to positive efficiencies are the machine $-+$ that employs process 2 flowing along its spontaneous tendency, to drive process 1 against its spontaneous tendency (e.g. heat engine) %in an elevator the spontaneous degradation of the electric potential into heat is used to increase the gravitational potential energy, thus performing work),
and the dual machine $+-$ where the system's spontaneous tendency is used to drive the environment against its tendency (e.g. the heat pump).}  
%(e.g. an elevator falling in a gravitational field could be employed to pump currents back into the power line).
Correspondingly we have $\theta(\eta) P_\tau(\eta) = P^{+-}_\tau(\eta) + P^{-+}_\tau(\eta)$ where
\bea
P_\tau^{+-}(\eta) = \int_{\substack{+ \sigma_1 > 0 \\ - \sigma_2 > 0}}  dx_1 dx_2 \, P_t(x_1,x_2) \, \delta\left(\eta + \frac{f_1 x_1}{f_2 x_2}\right)
 % \\ = \frac{e^{\frac{-\tau}{4} }}{2\pi \ALPHA \sqrt{|C|} } \left[1 - \sqrt{\pi \tau} \, \CHI \, e^{\tau \CHI^2} \mathrm{\,erfc}\,\left(+\sqrt{\tau} \CHI\right) \right], ~~
\eea
%yielding $P_t^{\pm \mp} = (2\pi \ALPHA \sqrt{|C|} e^{\frac{\tau}{4} })^{-1}\left[1 \mp \sqrt{\pi \tau} \, \CHI \, e^{\tau \CHI^2} \mathrm{\,erfc}\,\left(\pm \sqrt{\tau} \CHI\right) \right].$
and similarly for $P_\tau^{-+}$. Shaded plots are provided in Fig.\,\ref{distribution}, showing that each of the two maxima is almost exclusively determined by one of the two modes of the machine, the second of which by inversion of input/output has typical efficiency $1/\eta_m^\ast < 1$.  Regimes $++$ and $--$ contribute to the tail of the distribution at $\eta < 0$.

\begin{figure}
  \centering
 \includegraphics[width=240pt]{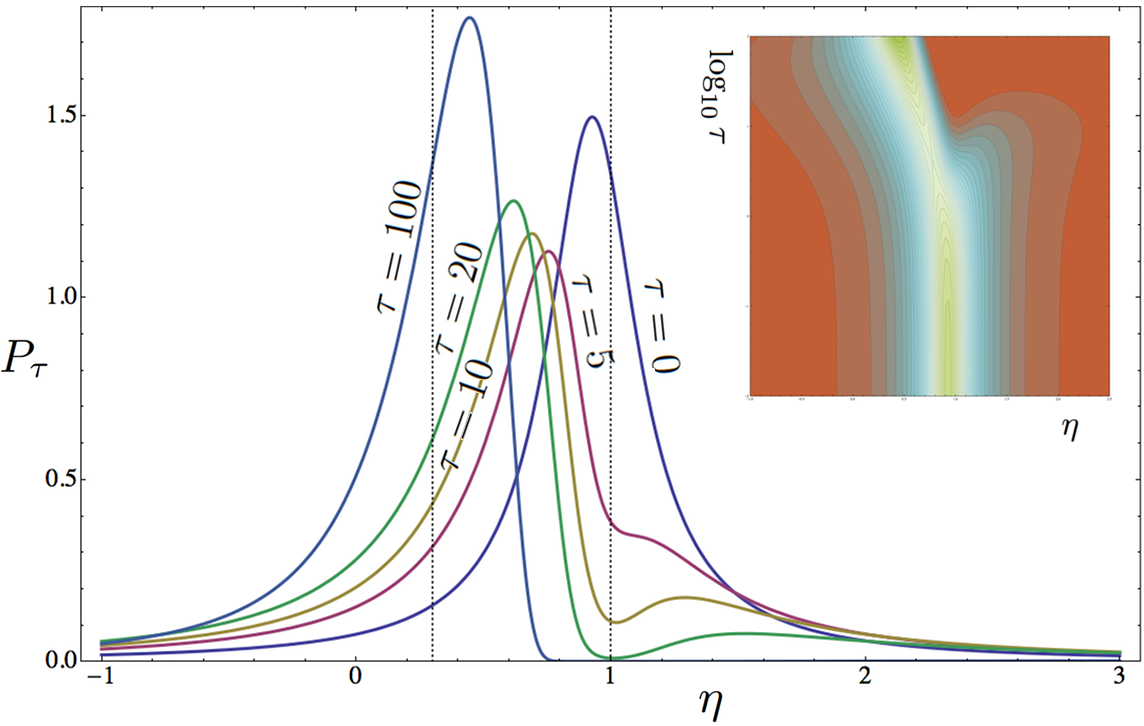}
  \caption{ \label{time} \tcb{Main frame: Efficiency distribution at various scaled times}, for  $\C = 0.05$, $\bar{\eta} = 0.3$, $\epsilon=+$.  The vertical dotted lines correspond to $\bar{\eta}$ and $\eta_c$. Inset: Contour plot of the efficiency p.d.f. as a function of $\eta$ and $\tau$ (in log scale). Maxima are points where  the level lines have horizontal tangents. After a critical time, a second maximum drifting to infinity appears.}
\end{figure}

Let us now study the  behavior of $P_\tau(\eta)$ in scaled time, depicted in Fig.\,\ref{time}.  At $\tau = 0$ we obtain a Cauchy distribution, 
$P_{0}(\eta) = 1/\big[\pi \ALPHA(\eta)  \sqrt{ |C|}\big]$, with maximum at $\eta_0 = -L_{12} f_1/(L_{22} f_2)$. 
We have $\eta_0 \geq \bar{\eta}$, and equality can only occur for $|C| = 0$. This implies that the most probable efficiency decreases in time \tc{towards $\bar{\eta}$}. Furthermore, at {\it thermodynamic equilibrium} where all the forces vanish, $\bs{f} \to 0$ at finite $\f$, it can be shown that $P^{\,\mathrm{eq.}}_\tau(\eta) = P_0(\eta)$, which means that systems at equilibrium do not evolve. 
% Interestingly, it can be shown that $P_0(\eta)$ is the probability distribution of a quantity that we call the {\it excess efficiency} $\eta_{\mathrm{exc}} = - (\sigma_1 - \bar{\sigma}_1)/(\sigma_2 - \bar{\sigma}_2)$.
As time elapses a transition to a bimodal distribution occurs, with the super-Carnot maximum drifting to infinity. We can define a critical time $\tau_{\mathrm{c}}$ at which there appears an inflection point in $P_\tau(\eta)$. %At large enough times the maxima of $P = P^{+-}+P^{-+}$ coincide with the  {\it plateaus} of $P^{+-}-P^{-+}$, which are determined by a quartic equation. Although analytical solutions of quartics are {\it de facto} impracticable, by this method we can prove the existence of $\tau_{\mathrm{c}}$.
\tcb{Numerical plots of $\tau_{\mathrm{c}}$ in terms of $\bar{\eta}$ and $c$ show that} the critical time is higher the closer to the maximal efficiency and to the {\it loose coupling} condition $\C\to 1$ \tcb{\cite{supp}}.
Finally, in the long time limit one has $\mathrm{\,erf}\,(\sqrt{\tau} \CHI) \sim 1 - e^{-\tau \CHI^2}/(\sqrt{\pi \tau} |\CHI|)$ \cite{sloane} and
\be
P_{\tau\to \infty}(\eta) \sim  \frac{ e^{- \frac{\tau}{4} } }{\pi \ALPHA(\eta)|C|} \left(1 - \frac{\CHI}{|\CHI|} + \sqrt{\pi\tau} e^{\tau \CHI(\eta)^2}  \right). \label{eq:larget}
\ee
The large-time behavior is captured by the large deviation rate function $I(\eta) = - \lim_{\tau \to \infty} \tau^{-1} \ln P_{\tau}(\eta) = 1/4 - \CHI(\eta)^2  \geq 0$, which was first calculated and thoroughly analyzed by Verley et al. \cite{gatien1,gatien2}. The rate function has only two extrema, a minimum $I(\bar{\eta}) = 0$ and a maximum $I(1) = 1/4$, and asymptotically $I(\pm \infty) = (4|C|(1-\bar{\eta})^2 + 4)^{-1} \leq I(1)$. Then, the more pronounced maximum tends to the macroscopic efficiency $\bar{\eta}$, while the minimum tends to the Carnot efficiency. The second maximum does not appear in the large deviation rate function  because at infinite time it moves to infinity, since it belongs to a subdominant decay mode.  This proves the existence of a critical time $\tau_{\mathrm{c}}$, as there must exist another maximum for the distribution to converge.

The quest for Carnot is very subtle. By Eq.\,(\ref{eq:limit}) the Carnot bound can be saturated in the limit $\C\to 0$, giving rise to two extreme situations \tcb{related to the spectrum and eigenvalues of the response matrix $L \to L^{\epsilon}$.} For $\epsilon = -$ ({\it tight coupling}), by Eq.\,(\ref{eq:C}) the correlation matrix becomes degenerate, 
\be
L^{-} = \left(\ba{cc} L_{11} & -\sqrt{L_{11} L_{22}} + O(\C) \\ - \sqrt{L_{11} L_{22} }  + O(\C) & L_{22} \ea\right), \label{eq:ltc}
\ee
where $O(\C)$ are terms of order $\C$. For $\epsilon = +$ ({\it singular coupling}), $L$   tends to the inverse of a degenerate matrix, i.e. $L^{+}=L^{-}/O(\C)$, with $|L^{+}| \to \infty$.

To reach Carnot, a second independent condition ({\it self-duality}) must hold:  $\f$ attains value  $\f^\ast = \sqrt{L_{11}/L_{22}}$, \tcb{which affords an interesting interpretation in terms of the probability of the inverse efficiency \cite{supp}. When  $\C\to 0$, this condition makes $\bs{f}$ either the null eigenvector of $L^-$ relative to its null eigenvalue, or of $L^+$ relative to its finite eigenvalue. In the tight-coupling regime, this condition is known as the {\it stall force} \cite{prost}.}
 
Expressing the efficiency in terms of the adimensional parameters $\zeta$ and $\phi = \varphi/\varphi^\ast$ (for $L_{12} < 0$) as \cite{entin}
\be
\bar{\eta}(\C,\phi) = - \frac{1-\phi \sqrt{1-\C}}{\phi^2-\phi \sqrt{1-\C}}\label{eq:etacv}
\ee
one finds that the two limits towards self-duality and towards tight/singular coupling do not commute,
\be
1 = \tc{-} \lim_{\C \to 0} \lim_{\phi \to 1} \bar{\eta} = + \lim_{\phi \to 1} \lim_{\C \to 0}  \bar{\eta}.
\ee
Then, a macroscopic Carnot efficiency is ``fragile'', as the self-dual forces needed to attain it are those that  slightly out of $\C=0$ give a  {\it dud}  machine that dissipates to obtain nothing, with macroscopic efficiency $\bar{\eta} = -1$.

% In particular, macroscopic self-dual machines are dud.  
\begin{figure}
  \centering
 \includegraphics[width=240pt]{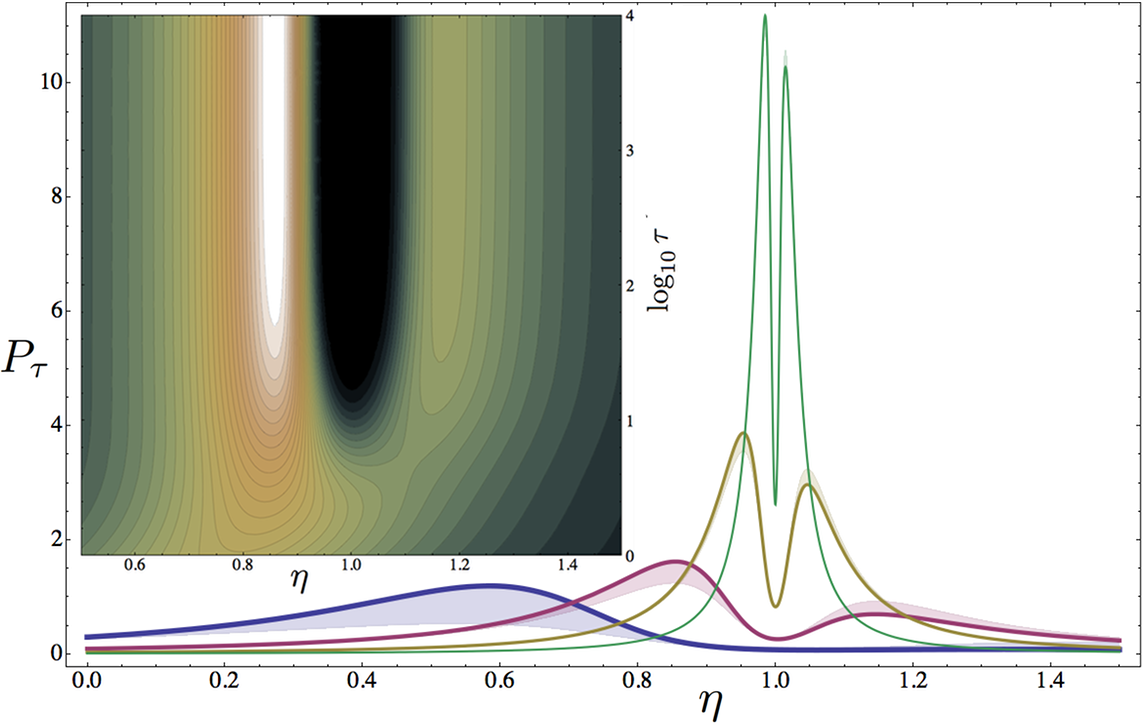}
  \caption{ \label{tight} Main: Graphs of $(\eta,P_\tau(\eta))$ for $\tau=10$, $\bar{\eta} = 0.3$, $\epsilon = +$ and for various coupling parameters (from bolder to thinner)  $\C = .1,.01,.001,.0001$. The shading represents the distance to the corresponding curves for $\bar{\eta}=-1$. Inset: Contour plot of the efficiency p.d.f.'s  corresponding to parameter $\C=0.1/\tau$ as a function of the efficiency $\eta$ and the scaled time $\tau$ (in log scale), \tcb{showing that the p.d.f. is invariant at all times, hence that singular coupling stretches the relaxation times.} Lighter tones for higher probabilities, darker for lower.}
\end{figure}

Nevertheless, the probabilistic level is richer. At tight coupling the bivariate Gaussian Eq.\,(\ref{eq:normal}) becomes univariate with support along $x_1/x_2 = - \f^\ast$, and the efficiency p.d.f. a Dirac delta centered at the macroscopic efficiency. 
% , $P_\tau(\eta) = \delta\left(\eta -  \bar{\eta}\right)$
Then tightly-coupled machines work macroscopically at all scaled times.

More interesting is the singular coupling.  Fig.\,\ref{tight} shows that in this limit all extrema tend to accumulate towards the Carnot efficiency, where the density concentrates. Despite the fact that the two peaks survive, convergence to a Dirac delta can be proven by the following argument \cite{argu}: From Eqs.\,(\ref{eq:pdf},\ref{eq:sub}), $\CHI \to 1/2$, $\ALPHA \to (1-\eta)^2$ and the efficiency p.d.f. converges to a distribution with support in $\eta =1$, which is then necessarily a finite combination of derivatives of the Dirac delta, $P_\tau(\eta) = \sum_{n=0}^N p_n \delta^{(n)}(1-\eta)$ \cite{hormander}. Since $\langle \,g\, \rangle > 0$ for all positive test functions $g(\eta) > 0$, then necessarily $p_{n} = 0$ but for $p_0 =1$ $\Box$.  Then, singular coupling pushes the most probable efficiency towards Carnot at fixed $\tau$; the shadings in Fig.\,\ref{tight} suggest that in this limit the distribution is fairly insensitive to $\bar{\eta}$. Moreover, the contour plot in Fig.\,\ref{tight} supports that the most probable efficiency stays at the same value for probability densities evaluated at a fixed time $\tau \propto 1/\C$, showing that convergence to $\bar{\eta}$ is more and more delayed. However, it must be remembered that the physical time scale is set by the entropy production rate. Necessarily the matrix entries of $L^{+}$ diverge; then in general $\bar{\sigma}$ also diverges. Still, $L^{+}$ admits a finite eigenvalue. Picking the forces along the relative eigenvector, $\phi = 1+ O(\C)$, one obtains a finite entropy production rate. Oddly, as discussed above, these conditions are met when the macroscopic machine is dud.

To resume: At singular coupling, the effect of fluctuations is macroscopically visible and permits to work close to Carnot efficiency at finite entropy production rate for sufficiently long physical times. The conditions for which the entropy production rate can be held finite are those under which the machine eventually evolves towards a dud fate. Notice that in this regime the system might flip randomly across the close sharp peaks of the p.d.f.. However, the inset in Fig.\,\ref{tight} suggests that at intermediate times reasonably high typical efficiencies will be favored and that a large separation between such peaks (the dark region of zero probability) occurs. Hence, to put it with a motto, a singular machine doomed to be  useless  might be efficiently useful for some time due to fluctuations; the better in the short run, the worse in the long. By the Green-Kubo relation Eq.\,(\ref{eq:gk}) the singular coupling limit is approached when correlations between the currents diverge and the inverse correlation matrix becomes degenerate. It is tempting to parallel this behavior to the paradigm of criticality at phase transitions, where fluctuations become macroscopic, correlations diverge and the covariance matrix degenerates \cite{zanardi,pole2}. \tcb{In practical applications, the best figure of merit reached is $zT \approx 3$ ($\C = 0.25$), in fact quite low. Then, we suggest that this insight might indicate a strategy to look for devices with higher figure of merit.} % At small scales, this could be of particular relevance for mechanisms like chemical switches and clocks that are believed to occur in  biological systems living close to bifurcation points \cite{switches}.

An important observation here to be made is that singular coupling pushes the system far from equilibrium. The framework of stochastic thermodynamics encompasses such systems by assuming that they are subtended by an underlying Markovian dynamics, giving rise to non-Gaussian current statistics. Gaussianity is only recovered in the linear regime at large times by the central limit theorem \cite{speck,engel}. \tcb{While the model of a Brownian particle in a tilted plane studied in Ref.\,\cite{gatien1} has exact Gaussian propagators as those studied in this paper, in general} Markov processes have a more complex behavior in time, in particular the average flux varies as the system evolves, depending on the initial ensemble. Then, the exact short- and large-time behavior of the efficiency distribution might become model-dependent. \tcb{For asymmetric protocols, a signature of non-Gaussian behavior is the off-Carnot least-probable efficiency \cite{gatien2,gingrich,roldan}.}

Nevertheless, our study points out that in the simplest Gaussian scenario the efficiency p.d.f. manifests peculiar features that might possibly be universal: Power-law tails, no finite moments, a naturally occurring transition to a bimodal distribution due to reverse working regimes, etc.  Particularly intriguing is the limit of a degenerate or singular covariance matrix. While the former case is intrinsically macroscopic  and broadly studied \cite{benenti, entin}, we obtain a clear indication that the singular coupling regime displays an interesting behavior that could lead to the enhancement of the efficiency above its macroscopic value. More light is to be shed on these issues by future inquiry on the finite-time statistics of the efficiency in stochastic models \cite{proesmans,prost} in their rich phenomenology, including maximum power generation \cite{espositomax,schmiedl}, multi-terminal machines \cite{mazza}, broken time-reversal symmetry \cite{saito}, the insurgence of phase transitions, and in relation to the issue of efficiency enhancement by noise \cite{hanggi} or by decoherence \cite{plenio}.
 \tcb{Experimental setups that could test these predictions are already available \cite{exp1,exp2,exp3,exp4,exp5}. The full statistics of the efficiency close to equilibrium has recently been sampled for a Carnot engine realized with a Brownian particle, in the quasistatic limit where the currents' statistics is Gaussian \cite{roldan}, and data analysis farther away from equilibrium might soon be available.}

\paragraph*{Aknowledgments.} The research was supported by the National Research Fund Luxembourg in the frame of project FNR/A11/02 and of Postdoc Grant 5856127.

\newpage

\section{Supplementary material}

\subsection{\label{app1}Efficiency p.d.f.}

In this section we derive the probability density function of the efficiency. Without possibility of confusion, we will denote stochastic variables by the values they take. Let us consider two  normally distributed stochastic variables $\bs{x} = (x_1,x_2)$ (the {\it fluxes})
\be
P_t(\bs{x}) = \frac{t}{4\pi \sqrt{|L|}} \exp \left[ - {\frac{t}{4} (\bs{x}-\bar{\bs{x}})\cdot  L^{-1} (\bs{x}-\bar{\bs{x}})} \right]. 
\ee
Defining $\bs{f} = L^{-1} \bar{\bs{x}}$ (the {\it forces}), we want to calculate the p.d.f. of the {\it efficiency}
\be
\eta = - \frac{f_1 x_1}{f_2 x_2}.
\ee
It is convenient to define  $\sigma_i = f_i x_i$ (the {\it entropy production rates}) with averages $\bar{\sigma}_i = f_i \bar{x}_i$. Let $\bar{\sigma} = \bar{\sigma}_1+\bar{\sigma}_2$. We have
\be
P'_t(\sigma_1,\sigma_2) d\sigma_1 d\sigma_2 = P_t(x_1,x_2) dx_1 dx_2
\ee
yielding
\be
P'_t(\sigma_1,\sigma_2) = \frac{t}{4\pi \bar{\sigma}\sqrt{|C|}} \exp \left[- {\frac{t}{4\bar{\sigma}} (\bs{\sigma}-\bar{\bs{\sigma}})\cdot  C^{-1} (\bs{\sigma}-\bar{\bs{\sigma}})}\right].
\ee
Letting $L_{ij}$ be the entries of $L$, then the matrix $C$ has entries $c_{ij} = L_{ij} f_if_j/\bar{\sigma}$ and it can be expressed as
\be
C
=\left(\ba{cc} \bar{\sigma}_1/\bar{\sigma} - c_{12} & c_{12} \\ c_{12} & \bar{\sigma}_2/\bar{\sigma} - c_{12}  \ea\right). 
\ee
Notice that  $P'_t$ only depends on three parameters, $\bar{\sigma}_1,\bar{\sigma}_2$ and $c_{12}$ or, equivalently, $\bar{\sigma}$, $\bar{\eta} = - \bar{\sigma}_1/\bar{\sigma}_2$ and $|C|$. Then the efficiency p.d.f. is given by
\bea
P_t(\eta) & = & \int d\sigma_1 d\sigma_2 \, \delta\left(\eta + \frac{\sigma_1}{\sigma_2}\right) P'_t(\sigma_1,\sigma_2) \nonumber \\ 
& = & \int d\sigma |\sigma| P'_t\left(-\eta \sigma,\sigma\right).
\eea
We have
\be
P'_t(- \eta \sigma,\sigma) = \frac{t}{4\pi \bar{\sigma}\sqrt{|C|}} \exp - {\frac{t}{4\bar{\sigma}} \left[ \ALPHA(\eta) \sigma^2 + 2\BETA(\eta) \sigma + \GAMMA \right]} \\
\ee
where we introduced
\bes
\ALPHA(\eta) & = & \frac{1}{|C|} (c_{22} \eta^2 + 2 c_{12} \eta + c_{11}) \\
\BETA(\eta) & = &  \frac{1}{|C|} [\eta (c_{22} \bar{\sigma}_1 - c_{12} \bar{\sigma}_2) + c_{12} \bar{\sigma}_1 - c_{11}\bar{\sigma}_2] \\
\GAMMA & = & \frac{1}{|C|}  (c_{22} \bar{\sigma}_1^2 - 2 c_{12} \bar{\sigma}_1 \bar{\sigma}_2 + c_{11} \bar{\sigma}_2^2).
\ees
Notice that $\ALPHA$ is adimensional, $\BETA$ has dimensions of an entropy rate, and $\GAMMA$ of a squared entropy rate. In fact, simple but tedious calculations show that
\bes
\ALPHA(\eta) & = & (1-\eta)^2+ \frac{1}{|C|}\left(\frac{\eta - \bar{\eta}}{1 - \bar{\eta}}\right)^2 \\
\BETA(\eta) & = & \bar{\sigma}(\eta-1) \\
\GAMMA & = & \bar{\sigma}^2 \phantom{\int}.
\ees
Notice that $\ALPHA \geq 0$ and equality can only hold if $|L| = 0$, a case that we hereby exclude. We now separate $P_t(\eta) = P^+_t(\eta) + P^-_t(\eta)$, where
\be
P^\pm_t (\eta) = \int_{0}^{\pm\infty} d\sigma \, \sigma P'_t(-\eta \sigma,\sigma).
\ee
Performing a change of variables $\sigma \to -\sigma$ in $P^-_t$, we have
\be
P^\pm_t = \frac{t}{4\pi\bar{\sigma} \sqrt{|C|}} \int_{0}^{\pm\infty} d\sigma \, \sigma  e^{- \frac{t}{4\bar{\sigma}} \left[ \ALPHA \sigma^2 \pm 2\BETA \sigma + \GAMMA \right]}. 
\ee
We can calculate the first and then change sign to $\BETA$ to obtain the second. We obtain
\bea
P^+_t & = & \frac{1}{2\pi \ALPHA\sqrt{|C|}} \int_{0}^{+\infty} d\sigma \, \frac{t (\ALPHA \sigma + \BETA - \BETA)}{2\bar{\sigma}} e^{- \frac{t}{4\bar{\sigma}} \left( \ALPHA \sigma^2 + 2\BETA \sigma + \GAMMA \right)} \nonumber \\
& = &   \frac{1}{2\pi \ALPHA \sqrt{|C|}}  \left\{ \int_{0}^{+\infty} d\sigma \left[ - \frac{d}{d\sigma} e^{- \frac{t}{4\bar{\sigma}} \left( \ALPHA \sigma^2 + 2\BETA \sigma + \GAMMA \right)} \right]  \right. \nonumber \\
& & \qquad \qquad  \qquad \qquad  \left. - \frac{t\BETA}{2\bar{\sigma}} \int_{0}^{+\infty} d\sigma \,e^{- \frac{t}{4\bar{\sigma}} \left( \ALPHA \sigma^2 + 2\BETA \sigma + \GAMMA \right)}  \right\} \nonumber \\
& = &   \frac{1}{2\pi \ALPHA  \sqrt{|C|}} e^{- \frac{t \GAMMA}{4\bar{\sigma}} }  \left\{1 - \frac{t\BETA}{2\bar{\sigma}} \int_{0}^{+\infty} d\sigma \,e^{- \frac{t}{4\bar{\sigma}} \left( \ALPHA \sigma^2 + 2\BETA \sigma \right)}  \right\} \nonumber \\
& = &   \frac{1}{2\pi \ALPHA  \sqrt{|C|}} e^{- \frac{t \GAMMA}{4\bar{\sigma}} }  \left\{1 -  \BETA e^{\frac{t  \BETA^2}{4\bar{\sigma} \ALPHA}}  \sqrt{\frac{t}{\bar{\sigma}\ALPHA}}  \int_{\BETA\sqrt{\frac{t}{4\bar{\sigma}\ALPHA}}}^{+\infty} d\sigma \,e^{- \sigma^2}  \right\}. \nonumber \\
\eea
Recognizing the complementary error function $\mathrm{\,erfc}(x) = 2/\sqrt{\pi} \int_x^{+\infty} e^{-y^2} dy$ and defining
\bea
\CHI(\eta) = \frac{-\BETA(\eta)}{2\bar{\sigma}\sqrt{\ALPHA(\eta)}}  =  \frac{1-\eta}{2\sqrt{\ALPHA(\eta)}}
\eea
we obtain 
\be
P^\pm_t = \frac{1}{2\pi \ALPHA \sqrt{|C|}} e^{- \frac{t\GAMMA}{4\bar{\sigma}} }  \left\{1 \pm \sqrt{\pi t\bar{\sigma}} \, \CHI \, e^{t\bar{\sigma} \CHI^2} \mathrm{\,erfc}\,(\mp \sqrt{t\bar{\sigma}} \CHI) \right\}.
\ee
Introducing $\tau = t \bar{\sigma}$, the full probability distribution reads
\be
P_t(\eta) = \frac{ e^{- \frac{\tau}{4} } }{\pi \ALPHA(\eta) \sqrt{|C|}} \left\{1 + \sqrt{\pi \tau} \, \CHI(\eta) \, e^{\tau \CHI(\eta)^2} \mathrm{\,erf}\,\left[\sqrt{\tau} \CHI(\eta)\right] \right\}
\ee
where we employed the fact that the error function $\mathrm{\,erf} = 1 - \mathrm{\,erfc}$ is odd.

Furthermore, defining
\bes
P_t^{++}(\eta) & = & \int_{\substack{\sigma_1 > 0 \\ \sigma_2 > 0}}  d\sigma_1 d\sigma_2 \, P'_t(\sigma_1,\sigma_2) \, \delta\left(\eta + \frac{\sigma_1}{\sigma_2}\right) \\
% & = & \theta(-\eta) \int_{0}^{+\infty} d\sigma \, \sigma P'_t(-\eta \sigma,\sigma) \\
P_t^{+-}(\eta) & = &  \int_{\substack{\sigma_1 > 0 \\ \sigma_2 < 0}}  d\sigma_1 d\sigma_2 \, P'_t(\sigma_1,\sigma_2) \, \delta\left(\eta + \frac{\sigma_1}{\sigma_2}\right) \\
% & = & \theta(+\eta) \int_{0}^{-\infty} d\sigma \, \sigma P'_t(-\eta \sigma,\sigma) \\
P_t^{-+}(\eta) & = & \int_{\substack{\sigma_1 < 0 \\ \sigma_2 > 0}}  d\sigma_1 d\sigma_2 \, P'_t(\sigma_1,\sigma_2) \, \delta\left(\eta + \frac{\sigma_1}{\sigma_2}\right) \\
% & = & \theta(+\eta) \int_{0}^{+\infty} d\sigma \, \sigma P'_t(-\eta \sigma,\sigma) \\
P_t^{--}(\eta) & = & \int_{\substack{\sigma_1 < 0 \\  \sigma_2 < 0}}  d\sigma_1 d\sigma_2 \, P'_t(\sigma_1,\sigma_2) \, \delta\left(\eta + \frac{\sigma_1}{\sigma_2}\right)  
% & = & \theta(-\eta) \int_{0}^{-\infty} d\sigma \, \sigma P'_t(-\eta \sigma,\sigma)
\ees
we obtain
\bes
P_t^{++}(\eta) & = & \theta(-\eta) P^+(\eta)  \\
P_t^{+-}(\eta) & = & \theta(+\eta) P^-(\eta) \\
P_t^{-+}(\eta) & = & \theta(+\eta) P^+(\eta)  \\
P_t^{--}(\eta) & = & \theta(-\eta)  P^-(\eta).
\ees

\subsection{Reparametrization}

The task is to express $|C|$ in terms of the parameters $\C = |C|/c_{11}c_{22}$ and $\bar{\eta}$. 
Notice that $c_{11} + 2c_{12} + c_{22}=1$. Then
\bes
\bar{\eta} & = & - \frac{c_{11}+c_{12}}{c_{22}+c_{12}} = - \frac{1 + c_{11}-c_{22}}{1 + c_{22}-c_{11}} \\
\C & = & \frac{|C|}{c_{11}c_{22}} = 1 - \frac{(1-c_{11}-c_{22})^2}{4(c_{11} c_{22})}.
\ees
From the first we obtain
\be
c_{22} = c_{11} + \frac{1+\bar{\eta}}{1-\bar{\eta}}
\ee
and letting $\xi = (1+\bar{\eta})/(1-\bar{\eta})$ from the second we get
\be
c_{11}^2 + \frac{\xi \C-1}{\C} c_{11} + \frac{(\xi-1)^2}{4\C} = 0
\ee
yielding
\bes
c_{11} & = & \frac{1 - \xi \C}{2\C} \left[1 \pm\sqrt{1 - \C \left(\frac{1-\xi}{1-\xi \C}\right)^2} \right] \nonumber \\
& = &  \frac{1 - \xi \C}{2\C}  \left(1 \pm \frac{1}{1 -  \bar{\eta} \frac{1+\C}{1-\C} } \sqrt{\bar{\eta}^2 - 2 \bar{\eta} \tfrac{1+\C}{1-\C} +1} \right)\qquad \\
c_{22} % & = & \frac{1 + \xi \C}{2\C} \left[1 \pm \sqrt{1 - \C \left( \frac{1+\xi}{1+\xi \C}\right)^2 } \right] \\
& = &  \frac{1 + \xi \C}{2\C} \left(1 \pm  \frac{1}{\frac{1+\C}{1-\C}  -  \bar{\eta} }
 \sqrt{\bar{\eta}^2 - 2 \bar{\eta} \tfrac{1+\C}{1-\C} +1}  \right).\qquad
\ees
Given
\bes
1 - \xi \C & = & \frac{1-\C}{1-\bar{\eta}} \left( 1 - \bar{\eta} \tfrac{1+\C}{1-\C} \right) \\
1+ \xi \C & = & \frac{1-\C}{1-\bar{\eta}} \left(  \tfrac{1+\C}{1-\C} - \bar{\eta} \right) 
\ees
we get
\bes
c_{11} & = & \frac{1-\C}{2\C(1-\bar{\eta})} \left( 1 - \bar{\eta} \tfrac{1+\C}{1-\C}   \pm \sqrt{\bar{\eta}^2 - 2 \bar{\eta} \tfrac{1+\C}{1-\C} +1} \right) \qquad \\
c_{22} & = & \frac{1-\C}{2\C(1-\bar{\eta})} \left( \tfrac{1+\C}{1-\C} - \bar{\eta} \pm \sqrt{\bar{\eta}^2 - 2 \bar{\eta} \tfrac{1+\C}{1-\C} +1}  \right). \qquad
\ees
We then obtain
\bea
|C| & = & \C \, c_{11}c_{22}\nonumber \\ 
& = &  \frac{(1-\C)^2}{4\C(1-\bar{\eta})^2} \left( 1 - \bar{\eta} \tfrac{1+\C}{1-\C}   \pm \sqrt{\bar{\eta}^2 - 2 \bar{\eta} \tfrac{1+\C}{1-\C} +1} \right) \times \nonumber \\
& & \qquad \times \left( \tfrac{1+\C}{1-\C} - \bar{\eta} \pm \sqrt{\bar{\eta}^2 - 2 \bar{\eta} \tfrac{1+\C}{1-\C} +1}  \right) \nonumber \\
% & = &  \frac{1-c}{2c (1-\bar{\eta})} \left(1 - \frac{1+c}{1-c} \bar{\eta} \right) \pm \frac{1-c}{2c}\sqrt{1 - \frac{4c}{1-c} \frac{\bar{\eta}}{(1-\bar{\eta})^2} }   \nonumber \\
& = & - \frac{\bar{\eta}}{(1-\bar{\eta})^2} + \frac{1-\C}{2\C} \left(1  \pm \sqrt{1 - \frac{4\C}{1-\C} \frac{\bar{\eta}}{(1-\bar{\eta})^2} } \right).  \nonumber \\
\eea

\subsection{Equilibrium case}

We consider the case $f_1,f_2 \to 0$ at fixed $\varphi = f_2/f_1$. We have to evaluate Eq.\,(2) in the main text
% \bea
% P_t(\eta) & = & \int dx_1 dx_2 \, \delta\left(\eta + \frac{x_1}{x_2 \varphi}\right) P_t(x_1,x_2) \nonumber \\ 
% & = & \varphi \int dx |x| P_t\left(-\eta \varphi \, x,x\right) 
% \eea
given that $\bs{x} = (x_1,x_2)$ is distributed with
\be
P_t(\bs{x}) = \frac{t}{\pi \sqrt{|L|}} \exp - {\frac{t}{4} \bs{x}\cdot  L^{-1} \bs{x}}.
\ee
We have
\be
P_t(- \eta \varphi x,x) = \frac{t}{\pi } \exp - {\frac{t}{4|L|} \left[ \ALPHA(\eta) x^2 + 2\BETA(\eta) x + \GAMMA \right]} \\
\ee
where we introduced
\be
\ALPHA(\eta) = \frac{1}{|L|}\left( L_{22} \varphi^2 \eta^2 + 2 L_{12} \varphi \eta + L_{11}\right).
\ee
All follows as for the derivation of the general efficiency p.d.f., but for $\BETA = \GAMMA = \CHI = 0$. Then the p.d.f. reads
\be
P_t(\eta) = \frac{2 \varphi }{\pi \ALPHA(\eta)\sqrt{|L|}}.
\ee
 
\subsection{Degenerate case (Tight coupling) \label{degenerate}}

Let $|L|=0$. Under our conditions $L_{12} \leq 0$, we have
\be
L_{12} = - \sqrt{L_{11} L_{22}}.
\ee
The eigenvectors of $L$ are: $(\sqrt{L_{11}},-\sqrt{L_{22}})^T$ relative to eigenvalue $L_{11} + L_{22} = \mathrm{\,tr}\,L$, and $(\sqrt{L_{22}},\sqrt{L_{11}})^T$ relative to eigenvalue $0$. Let
\be
U = \frac{1}{\sqrt{L_{11} + L_{22}}} \left(\ba{cc} \sqrt{L_{11}} & \sqrt{L_{22}} \\ - \sqrt{L_{22}} & \sqrt{L_{11}}  \ea \right) 
\ee
be the orthogonal matrix that performs the change of coordinate into the diagonal matrix $\Delta = \mathrm{\,diag}\,\{L_{11} + L_{22},0\}$. Then
\be
\sqrt{L} =  U \, \sqrt{\Delta} \, U^{T} = \frac{L}{\sqrt{L_{11}+L_{22}}}.
\ee
Furthermore, we introduce the Moore-Penrose pseudoinverse of $L$ which is obtained by inverting all nonvanishing eigenvalues:
\be
L^+  =  U \, \left(\ba{cc} (L_{11} + L_{22})^{-1} & 0 \\ 0 & 0 \ea \right) \, U^{T}  = \frac{L}{(L_{11}+L_{22})^2}
\ee
Letting $\Delta \bs{x} = (\bs{x}-\bar{\bs{x}})$, it is known that a degenerate normal distribution is supported along the direction
\be
\Delta \bs{x} = \sqrt{L} \,\bs{y} = \sqrt{\frac{L_{11} y_1 - L_{22} y_2}{L_{11}+L_{22}}} \left(\ba{c} \sqrt{L_{11}} \\ - \sqrt{L_{22}} \ea \right), \quad \bs{y} \in \mathbb{R}^2
\ee
and that it has probability density
\be
P_t(\bs{x}) = \delta\left(\Delta x_2 + \sqrt{\frac{L_{22}}{L_{11}}} \, \Delta x_1 \right)  \sqrt{\frac{t}{4 \pi L_{11}}} e^{- \frac{t}{4} \Delta \bs{x} \cdot L^+  \Delta \bs{x}}.
\ee
Moreover, the average currents are not arbitrary but we have $\bar{\bs{x}} = L \bs{f}$, which implies 
\be
P_t(\bs{x}) = \delta\left(x_2 + \sqrt{\frac{L_{22}}{L_{11}}} \, x_1 \right)  \sqrt{\frac{t}{4 \pi L_{11}}} e^{- \frac{t}{4} \Delta \bs{x} \cdot L^+  \Delta \bs{x}}.
\ee
Then, using a test function $g(\eta)$, we have
\bea
\langle \, g \, \rangle_t  % & = & \int d\eta \, g(\eta)  P_t(\eta)\nonumber \\
& = & \int d \eta \, g(\eta) \int dx_1 dx_2 \, \delta\left(\eta + \frac{x_1 f_1}{x_2f_2}\right) P_t(x_1,x_2) \nonumber \\ 
& = &   \int dx_1 dx_2 \,  g\left(- \frac{x_1}{x_2 \varphi }\right) \delta\left(x_2 + \sqrt{\frac{L_{22}}{L_{11}}} \, x_1 \right) \times \nonumber \\
& & \qquad \times \, \sqrt{\frac{t}{4 \pi L_{11}}} \, e^{- \frac{t}{4} \Delta \bs{x} \cdot L^+  \Delta \bs{x}}
 \nonumber \\
& = &  g\left(\varphi^\ast/\varphi\right).
\eea
By Eq.\,(16) in the main text at $\C \to 0$, $\varphi^\ast/\varphi \to \bar{\eta}$.

\subsection{Fluctuation relation for self-dual p.d.f.}

The two maxima of $P_\tau(\eta)$ are due to converse regimes. It is then natural to define the stochastic variable $\eta^\ast = -\sigma_2/\sigma_1$ and look at its p.d.f.
\bea
P^\ast_\tau(\eta) = \eta^{-2}P_\tau(1/\eta)
\eea
Self-duality is the condition upon which $P^\ast_\tau(\eta) = P_\tau(\eta)$, implying the following fluctuation relation for the efficiency
\bea
P^{\mathrm{\,s.d.}}_\tau(\eta) = \frac{1}{\eta^2} \, P^{\mathrm{\,s.d.}}_\tau(1/\eta)
\eea

Let us show that the choice $\varphi = \varphi^\ast$ yields this fluctuation relation. First, it simple to show that $\varphi = \varphi^\ast = \sqrt{L_{11}/L_{22}}$  implies the symmetry of the normal multivariate for the currents
\be
P^{\mathrm{\,\mathrm{s.d.}}}_t(x_1,x_2) = P^{\mathrm{\,\mathrm{s.d.}}}_t\left(\varphi x_2,{\varphi}^{-1} x_1\right). \label{eq:sym}
\ee
In fact, the Jacobian of the transformation $(x_1,x_2) \to (\varphi x_2,\varphi ^{-1} x_1)$ is $1$, and the identity between the quadratic polynomials at exponent can be checked by direct substitution:
\begin{multline}
L_{22} (x_1 - \bar{x}_1)^2 - 2 L_{12} (x_1 - \bar{x}_1) (x_2 - \bar{x}_2) +  L_{11} (x_2 - \bar{x}_2)^2 =  \\
= L_{22} (\varphi  x_2 - \bar{x}_1)^2 - 2 L_{12} (\varphi  x_2 - \bar{x}_1) (x_1/\varphi  - \bar{x}_2) +  L_{11} (x_1/\varphi  - \bar{x}_1)^2
\end{multline}
In fact, equating order by order one can also show that the choice $\varphi = \pm \varphi^\ast$ are the only ones yielding Eq.\,(\ref{eq:sym}).

Now let us consider the efficiency p.d.f.:
\bea
P^{\mathrm{\,s.d.}}_t(\eta) & = & \varphi \int dx |x| P^{\mathrm{\,s.d.}}_t\left(-\eta \varphi x,x\right) \\
& = &  \varphi \int dx |x| P^{\mathrm{\,s.d.}}_t\left(\varphi x,- \eta x\right) \nonumber \\
& = &  \frac{\varphi}{|\eta|\eta} \int_{y(-\infty)}^{y(+\infty)} dy |y| P^{\mathrm{\,s.d.}}_t\left(-\varphi y/\eta,y\right) \nonumber \\
& = &  \frac{\varphi}{\eta^2} \int dy |y| P^{\mathrm{\,s.d.}}_t\left(-\varphi y/\eta,y\right)
~=~  \frac{1}{\eta^2} P^{s.d.}_t(1/\eta) \nonumber
\eea
where on the third line we performed the change of coordinates $y(x) = - \eta x$ and we kept into account the order of the extremes of integration by including a suitable absolute value.

\subsection{Efficiency at singular coupling}

We  say the covariance matrix tends to become singular when its inverse tends to become degenerate. Let $\varepsilon(\C)$ be small of order $\C$. An almost degenerate inverse takes the form
\bea
L^{-1} = \left(\ba{cc} m_{11} & (1-\varepsilon) \sqrt{m_{11} m_{22}} \\ (1-\varepsilon) \sqrt{m_{11} m_{22} } & m_{22} \ea\right).
\eea
Then we have
\bea
L = \frac{1}{2 \varepsilon} \left(\ba{cc} L_{11} & (\varepsilon-1) \sqrt{L_{11} L_{22}} \\ (\varepsilon-1) \sqrt{L_{11} L_{22}} & L_{22} \ea\right)
\eea
where $L_{11}=m_{11}^{-1}$, $L_{22}=m_{22}^{-1}$. Eigenvalues:
\bes
\lambda_+ & = & \frac{1}{2\epsilon}(L_{11} + L_{22}) \left(1 - 2\varepsilon \frac{L_1L_2}{ (L_{11} + L_{22})^2} \right) \\ 
\lambda_- & = & \frac{L_1L_2}{L_{11} + L_{22}} . 
\ees
The first is divergent, the second finite. The eigenvector $\bs{f}_-$ relative to the finite eigenvalue is such that
\be
\frac{f_1}{f_2} = \sqrt{\frac{L_{22}}{L_{11}}}\left(1+\varepsilon \frac{L_{11}-L_{22}}{L_{11}+L_{22}}\right).
\ee
Therefore
\be
\phi = \varphi/\varphi^\ast = \left(1+\varepsilon  \frac{L_{22}-L_{11}}{L_{11}+L_{22}}\right).
\ee
By Eq.\,(16) the macroscopic efficiency along this eigenvectors reads
\be
\bar{\eta} = -1 + O(\C^2).
\ee

\subsection{Critical time}

The critical time is defined as the scaled time at which an inflection point appears in the p.d.f. of the efficiency. Fig.\,\ref{critical} provides a color plot of the critical time.

\begin{figure}[h!]
  \centering
 \includegraphics[width=240pt]{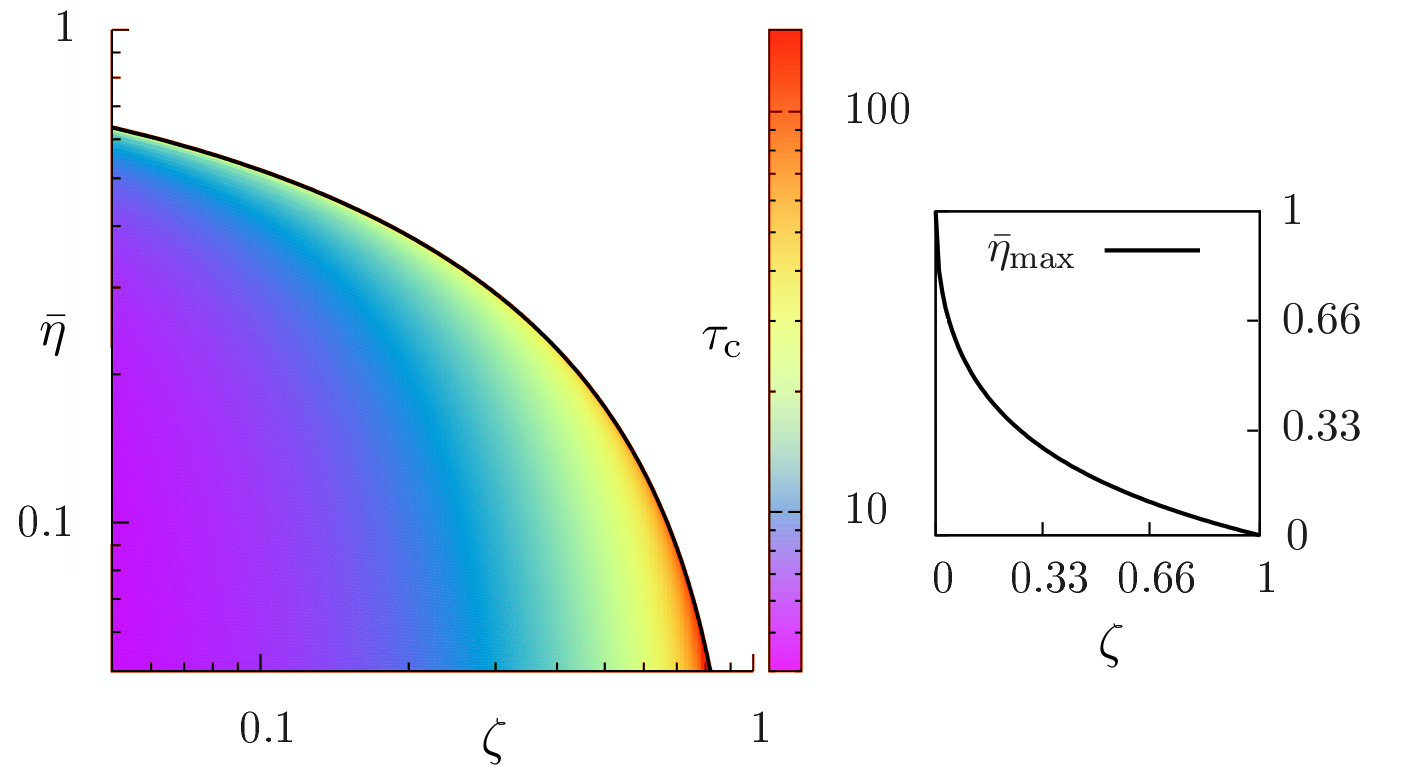}
  \caption{ \label{critical}Color plot of $\tau_{\mathrm{c}}$ as a function of the tight coupling parameter $\C$ and of the macroscopic efficiency $\bar \eta$, for $\epsilon = +$, \tc{in log-log scale. Inset: maximal efficiency} as a function of the coupling parameter, in natural scale.}
\end{figure}

\end{document}